\documentclass[english,prd,nofootinbib,preprintnumbers,
reprint,
 amsmath,amssymb,
 aps,
]{revtex4-1}

\usepackage{graphicx}
\usepackage{dcolumn}
\usepackage{bm}

\usepackage{color}
\input{colordvi.tex}

\newcommand{\beqa}{\begin{eqnarray}}
\newcommand{\eeqa}{\end{eqnarray}}
\newcommand{\simg}{\gtrsim}

\begin{document}

\preprint{APS/123-QED}

\title{Planck constraints on scalar-tensor cosmology \\ 
and the variation of the gravitational constant}

\author{Junpei Ooba${}^1$}
\author{Kiyotomo Ichiki${}^{1,2}$}
\author{Takeshi Chiba${}^3$}
\author{Naoshi Sugiyama${}^{1,2,4}$}

\affiliation{%
	${}^1$Department of physics and astrophysics, Nagoya University, Nagoya 464-8602, Japan\\ 
	${}^2$Kobayashi-Maskawa Institute for the Origin of Particles and the Universe, Nagoya University, Nagoya 464-8602, Japan\\
	${}^3$Department of Physics, College of Humanities and Sciences, Nihon University, Tokyo 156-8550, Japan\\
	${}^4$Kavli Institute for the Physics and Mathematics of the Universe (Kavli IPMU),\\
	The University of Tokyo, Chiba 277-8582, Japan
}%



\date{\today}

\begin{abstract}
Cosmological constraints on the scalar-tensor theory of gravity
by analyzing the angular power spectrum data of the cosmic microwave background (CMB) obtained from the Planck 2015 results
are presented.
We consider the harmonic attractor model,
in which the scalar field has a harmonic potential with curvature ($\beta$) in the Einstein frame and the theory relaxes toward the Einstein gravity with time.
Analyzing the {\it TT}, {\it EE}, {\it TE} and lensing CMB data from Planck by the Markov chain Monte Carlo method,
we find that the present-day deviation from the Einstein gravity (${\alpha_0}^2$) is constrained
as ${\alpha_0}^2<2.5\times10^{-4-4.5\beta^2}\ (95.45\% {\rm\ C.L.})$ and ${\alpha_0}^2<6.3\times10^{-4-4.5\beta^2}\ (99.99\%\ {\rm C.L.})$ for $0<\beta<0.4$.
The time variation of the effective gravitational constant between the recombination and the present epochs is constrained as
$G_{\rm rec}/G_0<1.0056\ (95.45\%  {\rm\ C.L.})$ and $G_{\rm rec}/G_0<1.0115\ (99.99 \%{\rm\ C.L.})$.
We also find that the constraints are little affected  by extending to nonflat cosmological models
because the diffusion damping effect revealed by Planck breaks the degeneracy of the projection effect.

\begin{description}
\item[PACS numbers]
04.80.Cc, 98.80.Es
\end{description}
\end{abstract}
\pacs{Valid PACS appear here}
\maketitle


\section{\label{sec:level1}Introduction}

Unifying the elementary forces \cite{string} is among the most important goals of modern 
physics.  One proposition motivated by superstring theory, which is the most plausible 
 candidate of the unified theory including gravity, is that the physical constants are 
affected by the vacuum expectation values (VEVs) of scalar fields.
Accordingly, it is natural to consider time variation of these physical constants as the VEVs of 
scalar fields (such as string dilaton) vary (see \cite{chiba} for the experimental constraints on the time variation of physical constants). 
The dilaton gravity is 
classified as one of the scalar-tensor theories of gravity.  
In the scalar-tensor theories of gravity, a scalar field couples to 
the Ricci scalar, which provides a natural framework for realizing the time variation of 
the gravitational constant via the dynamics of the scalar field. 
In the Jordan-Brans-Dicke theory of gravity \cite{BDT}, which is the simplest example of 
 scalar-tensor theories, a constant coupling parameter $\omega$ is introduced.
In more general scalar-tensor theories \cite{st}, $\omega$ is promoted to a function of the 
 Brans-Dicke scalar field $\phi$.
In the limit $\omega \rightarrow \infty$, the Einstein gravity is recovered and the gravitational constant 
becomes a constant in time.  

The coupling parameter $\omega$ has been constrained by  several solar system experiments.
For instance, the weak-field experiment conducted in the Solar System by the Cassini mission has put strong constraints on the post-Newtonian deviation from the Einstein gravity, where $\omega$ is constrained as $\omega>43000$ at $2\sigma$ level \cite{Cassini,Will}.

For the cosmological scale experiments, the possibility of constraining the Brans-Dicke theory by temperature and polarization anisotropies of the cosmic microwave background (CMB) was suggested in \cite{ck}, and  Nagata {\it et al.} \cite{NCS2004} first placed constraints on a general scalar-tensor theory called the harmonic attractor model including the Jordan-Brans-Dicke theory \cite{dn}.
In this model the scalar field has a quadratic effective potential of positive curvature in the Einstein frame, and the Einstein gravity is an attractor that naturally suppresses any deviations from the Einstein gravity in the present epoch.
Nagata {\it et al.} reported that the present-day value of $\omega$ is constrained as $\omega>1000$ at $2\sigma$ level by analyzing the CMB data from the Wilkinson Microwave Anisotropy Probe (WMAP). Moreover,  the gravitational constant at 
the recombination epoch $G_{\rm rec}$ relative to the present gravitational 
constant $G_0$ is constrained as $G_{\rm rec}/G_0<1.05$ ($2\sigma$). These constraints basically come from the fact that the size of the sound horizon at the recombination epoch, which determines the characteristic angular scale in the angular power spectrum of CMB anisotropies, depends on the amounts of matter and baryon contents and on the strength of the gravity at that epoch.
The time variation of the gravitational constant indeed shifts the locations and the amplitudes of acoustic peaks in the CMB angular power spectrum.
In some parameter regions of the harmonic attractor model, the constraints become stronger 
 than those found in the Solar System analyses.

In this paper, we further constrain the parameter $\omega$ and the time variation of the gravitational constant  in the  harmonic attractor model by analyzing the latest CMB temperature and polarization anisotropy spectra from Planck \cite{Planck}. 
We also investigate how large the time variation of the gravitational constant is allowed by comparing its values at the recombination and present epochs.
Avilez and Skordis \cite{Avilez} placed a constraint on $\omega$ as $>890$ at $99\%$ confidence level (C.L.) by analyzing the CMB data from Planck 2013.
Ballardini {\it et al.} \cite{Ballardini}  studied the 
constraints from Planck 2015 data on the induced gravity 
dark energy model with a quartic potential 
 which can be cast into a Jordan-Brans-Dicke model with a quadratic potential 
and reported  the constraints on $\omega$ and on  the gravitational constant 
at the radiation epoch  $G_{\rm rad}$ as $\omega>147$ and $G_{\rm rad}/G_0<1.039$ at $95\%$ C.L., respectively (see also \cite{Umilta} for the Planck 2013 data).

The remainder of the paper is organized as follows: Sec. II explains the scalar-tensor cosmological model and the changes in the angular power spectrum of the CMB temperature anisotropy.
Section III describes our method for constraining the scalar-tensor coupling parameters.
In Sec. IV, we compare the model with the Planck data.  Finally, our conclusions are presented in Sec. V.

Unless stated otherwise, numerical calculations performed for illustration purpose assume the standard values of the cosmological parameters:
$h=0.67556$,
$\Omega_b h^2=0.022032$, $\Omega_c h^2=0.12038$, $z_{\rm re}=11.357$,
$A_s=2.215\times 10^{-9}$, $n_s=0.9619$, where $h$ is the Hubble
parameter, $\Omega_b h^2$ and $\Omega_c h^2$ are the density parameters for
baryon and cold dark matter components, respectively, $z_{\rm re}$ is
the reionization redshift, and $A_s$ and $n_s$ are the amplitude and
spectral index of primordial curvature fluctuations, respectively. 

\section{Mathematics and Equations}

We briefly review the cosmological background and perturbation equations that are given in Nagata {\it et al.} \cite{NCS2002}.

The action describing a general massless scalar-tensor theory in the 
Jordan frame is given by
\begin{equation}
S =\frac{1}{16\pi G_0}\int d^4x\sqrt{-g}\left[ \phi R - \frac{\omega(\phi)}{\phi}(\nabla \phi)^2 \right] + S_{\rm m}[\psi,g_{\mu\nu}],
\label{eq:action}
\end{equation}
where $G_0$ is the present-day Newtonian gravitational constant,
$S_{\rm m}[\psi,g_{\mu\nu}]$ is the matter action which is a function of the matter variable $\psi$ and the metric $g_{\mu\nu}$. 
This ``Jordan frame metric'' defines the lengths and times
actually measured by laboratory rods and clocks since, in the action Eq.(\ref{eq:action}), matter is universally coupled to $g_{\mu\nu}$  \cite{de,cy}.
The function $\omega(\phi)$ is the dimensionless coupling parameter which depends on the scalar 
field $\phi$. We set $\omega(\phi)$ to the following form,
\begin{equation}
2\,\omega(\phi) + 3 = \left\{ {\alpha_0}^2 - \beta\, {\rm ln}(\phi/\phi_0) \right\}^{-1},
\label{eq:omega}
\end{equation}
where $\phi_0$, $\alpha_0$ and $\beta$ are the present values of the $\phi$, potential gradient and curvature, respectively.

The background equations for a Friedmann universe are
\begin{equation}
\rho' = -3\frac{a'}{a}(\rho + p),
\label{eq:EoS}
\end{equation}
\begin{equation}
\left( \frac{a'}{a} \right)^2 + K= \frac{8\pi G_0\,\rho\, a^2}{3\, \phi} - \frac{a'}{a}\frac{\phi'}{\phi} + \frac{\omega}{6}\left( \frac{\phi'}{\phi} \right)^2,
\label{eq:friedmann}
\end{equation}
\begin{equation}
\phi'' + 2\frac{a'}{a}\phi' = \frac{1}{2\, \omega + 3}\left\{ 8\pi G_0\,a^2(\rho - 3p) - {\phi'}^2\frac{d\omega}{d\phi} \right\},
\label{eq:phiEoM}
\end{equation}
where $a$ is the cosmological scale factor and the prime notation denotes a derivative with respect to the conformal time,
$\rho$ and $p$ are the total energy density and pressure, respectively,
and $K$ denotes a constant spatial curvature.

The effective gravitational constant measured by Cavendish-type experiments is given by \cite{de}
\begin{equation}
G(\phi) = \frac{G_0}{\phi}\frac{2\, \omega(\phi) + 4}{2\, \omega(\phi) + 3}.
\label{eq:Gphi}
\end{equation}
The present value of $\phi$ must yield the present-day Newtonian gravitational constant and satisfy the expression of $G(\phi_0) = G_0$.
Thus, we have
\begin{equation}
\phi_0 = \frac{2\, \omega_0 + 4}{2\, \omega_0 + 3} = 1 + {\alpha_0}^2,
\label{eq:phi0}
\end{equation}
where $\omega_0$ is the present value of $\omega(\phi)$.

\begin{figure}[ht]
\includegraphics[width=9cm,]{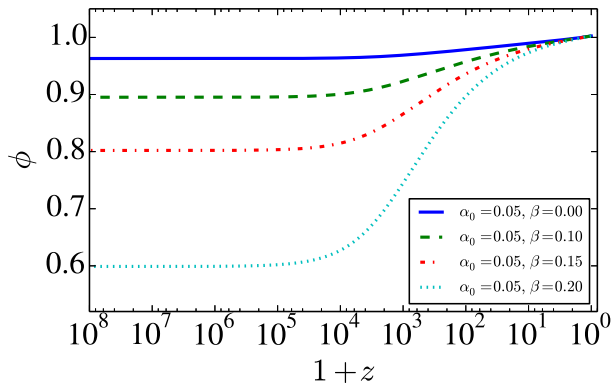}
\caption{\label{fig:BDphi} Time evolution of $\phi$ in the scalar-tensor
 $\Lambda {\rm CDM}$ model, with the parameters as indicated in the
 figure. The other cosmological parameters are fixed to the standard values.}
\end{figure}

\begin{figure}[ht]
\includegraphics[width=9cm,]{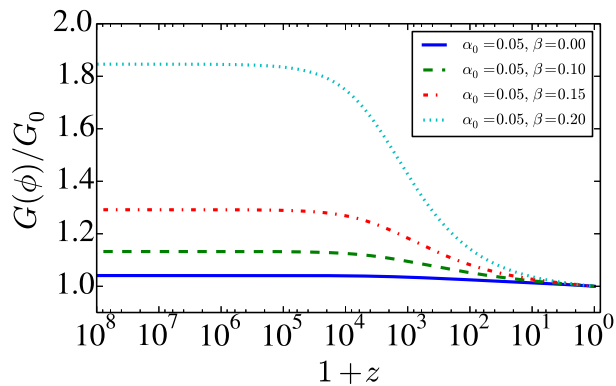}
\caption{\label{fig:Geff} Time evolution of $G(\phi)/G_0$ in the
 scalar-tensor models with the same parameters as in Fig.~\ref{fig:BDphi}.
The effective gravitational constant $G(\phi)$ is inversely proportional to the scalar field $\phi$ through Eq. (\ref{eq:Gphi}).}
\end{figure}

Typical evolutions of $\phi$ and $G(\phi)$ are shown in Figs. \ref{fig:BDphi} and \ref{fig:Geff}, respectively.
In the radiation-dominated epoch, $\phi$ stays constant because the pressure of the relativistic component in Eq. (\ref{eq:phiEoM}) is $p=\rho/3$.
As the universe evolves toward matter-radiation equality, $\phi$ begins growing and finally converges at $\phi_0$,
realizing the present-day Newtonian gravitational constant.
During the evolution to the present epoch, gravity deviates slightly and smoothly from the Einstein gravity.

\begin{figure}[ht]
\includegraphics[width=9cm,]{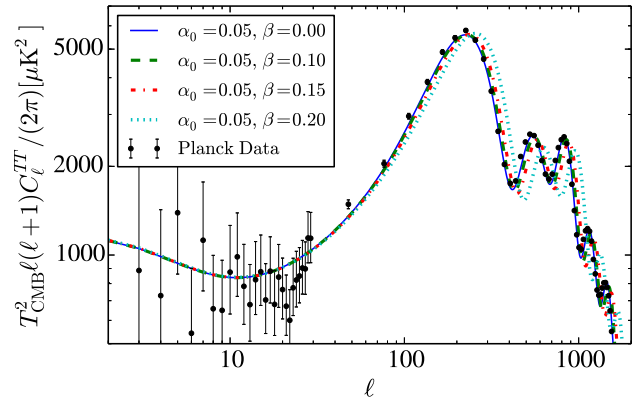}
\caption{\label{fig:errorCl} CMB temperature anisotropy spectra in the scalar-tensor models with the $\Lambda {\rm CDM}$ parameters.
The data points with error bars represent the Planck data.}
\end{figure}

The variation in the value of $\phi$ alters the Hubble parameter in the early universe
from its value under the Einstein gravity through Eq. (\ref{eq:friedmann}).
As shown in Fig. \ref{fig:Geff}, the harmonic attractor model always predicts a larger gravitational constant in the early universe as long as $\alpha_0^2 $ is non-negative
and hence a smaller horizon length at a given redshift.
Because the locations of the acoustic peaks and the damping scale depend differently on the horizon length at recombination,
we can constrain the $\phi$-induced variations in the horizon scale by analyzing the precisely measured CMB anisotropies on small angular scales.
The shift of the acoustic peaks to smaller angular scales is proportional to the horizon length ($\propto H^{-1}$),
while that of the damping scale is less affected by it ($\propto \sqrt{H^{-1}}$).
Therefore, the first peak and the diffusion tail in the angular power spectrum become closer as the expansion rate becomes larger,
suppressing the small scale peaks, as shown in Fig. \ref{fig:errorCl}. 


\section{METHODS}

To compute the temperature and polarization fluctuations in the CMB and the lensing potential power spectra, we
numerically solve the equations in the model described in the previous
section modifying the publicly available numerical code, CLASS \cite{CLASS2}.  The data are 
 analyzed using
the Markov chain Monte Carlo (MCMC) method with Monte Python \cite{Audren}
developed in the CLASS code.
In our calculations, we consider ($\alpha_0$,\ $\beta$) in Eq. (\ref{eq:omega}), which characterize the scalar-tensor theory,
in addition to the parameters of the $\Lambda {\rm CDM}$ model.

We set the priors for the standard cosmological parameters as
\begin{align}
\label{eq:prior}
H_0 \in (30,100),\ \ \Omega_{\rm b}h^2 \in (0.005,0.04), \nonumber\\
\Omega_{\rm c}h^2 \in (0.01,0.5),\ \ \tau_{\rm reio} \in (0.005,0.5), \\
{\rm ln}(10^{10}A_{\rm s}) \in (0.5,10),\ \ n_{\rm s} \in (0.5,1.5), \nonumber
\end{align}
and for $\alpha_0$ and $\beta$ as
\begin{align}
\label{eq:prior2}
\alpha_0 &\in (0,0.5), \\
\beta &\in (0,0.4).
\label{eq:prior3}
\end{align}
The CMB temperature and the effective number of neutrinos were set
to $T_{\rm CMB}= 2.7255\ \rm K$ from COBE \cite{Fixsen} and $N_{\rm
eff}=3.046$, respectively.
The primordial helium fraction $Y_{\rm He}$ is inferred from the standard Big Bang nucleosynthesis, as a function of the baryon density \cite{BBN}.
We compare our results with the CMB
angular power spectrum data from the Planck 2015 mission \cite{Planck},
which include the auto power spectra of temperature and polarization anisotropies ({\it TT} and {\it EE}),
their cross-power spectrum ({\it TE}), and the lensing potential power spectrum.

Because the variation of the gravitational constant could alter the distance to the last scattering surface of the CMB,
its effect on the angular power spectrum may degenerate with the effect of spatial curvature in the Friedmann universe.
Therefore, we separately perform a MCMC analysis for models with the spatial curvature ($\Omega_{\rm K}$). We set a prior for $\Omega_{\rm K}$ as
\begin{align}
\Omega_{\rm K} \in (-0.5,0.5),
\end{align}
while the same priors are used for the other standard cosmological
parameters and ($\alpha_0$,\ $\beta$) as shown in Eqs. (\ref{eq:prior}), (\ref{eq:prior2}) and (\ref{eq:prior3}).

\section{\label{sec:level1}RESULTS}

In this section, we show the results of the parameter constraints.

\subsection{\label{sec:level2}Flat universe case} 

\begin{figure}[ht] 
\includegraphics[width=7cm,]{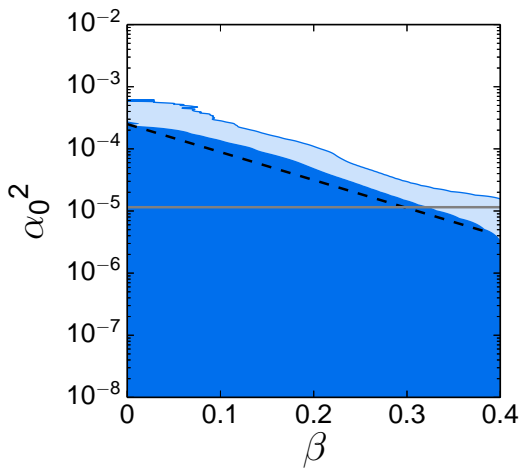}
\caption{\label{fig:contour} $95.45\%$ and $95.45\%$ confidence contours in the ${\alpha_0}^2-\beta$ plane
for the scalar-tensor $\Lambda {\rm CDM}$ models with the other parameters marginalized.
The black dashed line shows the function ${\alpha_0}^2 = 2.5 \times 10^{-4-4.5\beta}$
and the gray solid line shows the bound from the Solar System experiment.
}
\end{figure}

In Fig. \ref{fig:contour}, we show the constraint contours in the ${\alpha_0}^2-\beta$ plane, where the other parameters are marginalized.
We find that the scalar-tensor coupling parameters are constrained as
\begin{align}
{\alpha_0}^2 &< 2.5 \times 10^{-4-4.5\beta}\ \ (95.45\%), \\
{\alpha_0}^2 &< 6.3 \times 10^{-4-4.5\beta}\ \ (99.99\%),
\end{align}
where the number in the parenthesis denotes the confidence level. 
The change of the scalar field from the CMB epoch to the present is larger
for either a larger $\alpha_0^2$  model or a larger $\beta$  model (see  Fig. 1 and Fig. 2 in \cite{NCS2002}), which induces the degeneracy.
This result can be translated into the present-day value of the coupling parameter $\omega$ at $\beta = 0$ using Eq. (\ref{eq:omega}) as
\begin{align}
\omega &> 2000\ \ (95.45\%), \\
\omega &> 790\ \ (99.99\%).
\end{align}

Previously,
Nagata {\it et al.} \cite{NCS2004} reported that ${\alpha_0}^2 < 5 \times 10^{-4-7\beta}\ (10^{-2-7\beta})$ at 2$\sigma$ (4$\sigma$) level,
which corresponds to $\omega > 1000\ (50)$ at 2$\sigma$ (4$\sigma$) level.
Our constraints are significantly improved over this value. Our results are 
complementary to those by Avilez and Skordis \cite{Avilez} who 
reported $\omega>890$ at 99\% C.L. for the constant $\omega$ model
(our result is $\omega>1100$ at 99\% C.L.).
Furthermore, in the large $\beta$ regime ($\beta \hspace{0.3em}\raisebox{0.3ex}{$>$}\hspace{-0.8em}\raisebox{-.8ex}{$\sim$}\hspace{0.2em}0.3$),
our cosmological constraint is stronger than that determined in the Solar System study
($\omega>43000$, which corresponds to ${\alpha_0}^2 < 1.15 \times 10^{-5}$) \cite{Will,Cassini}.

Table \ref{tab:table1} shows $68.27\%$ confidence limits of the standard cosmological parameters in the scalar-tensor $\Lambda \rm CDM$ model.
These parameters are still consistent with those of the Planck results \cite{Planck} in the standard $\Lambda {\rm CDM}$ model.
Table \ref{tab:table2} shows $95.45\%$ confidence limits of the parameters ${\rm log}_{10}({\alpha_0}^2)$ and $\beta$.\\

Next, we consider the variation of the gravitational constant in the recombination epoch.
We define $G_{\rm rec} \equiv G(\phi_{\rm rec})$ and put constraints on $G_{\rm rec}/G_0$,
after marginalizing over the other parameters.
Here, $\phi_{\rm rec}$ is the value of $\phi$ at the recombination epoch when the visibility function takes its maximum value. 
We compute the  marginalized posterior distribution of $G_{\rm rec}/G_0$
as shown in Fig. 5 (for flat models).   
We find that $G_{\rm rec}/G_0$ is constrained as
\begin{align}
\label{eq:Gbound1}
G_{\rm rec}/G_0 &< 1.0056\ \ (95.45\%), \\
G_{\rm rec}/G_0 &< 1.0115\ \ (99.99\%).
\label{eq:Gbound2}
\end{align}

\begin{figure}[ht]
\includegraphics[width=7cm,]{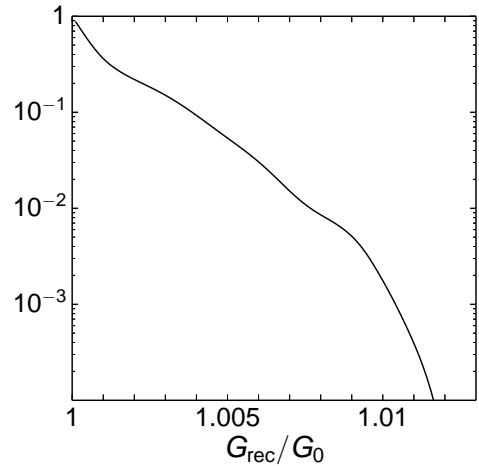}
\caption{\label{fig:Gphi_1d} Posterior distribution of $G_{\rm rec}/G_0$.
}
\end{figure}

\begin{figure}[ht]
\includegraphics[width=9cm,]{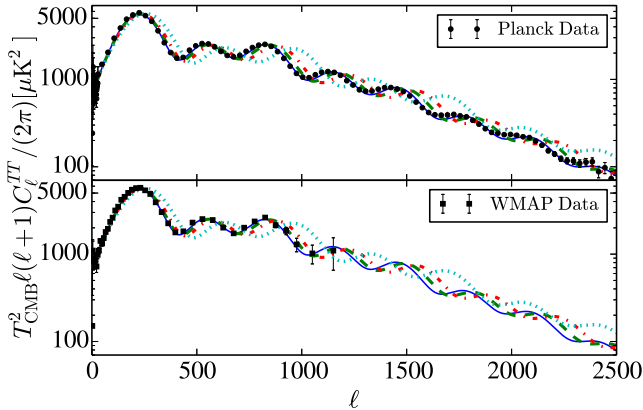}
\caption{\label{fig:compare} Comparison of the CMB temperature anisotropy spectra between Planck and WMAP.
CMB temperature anisotropy spectra in the scalar-tensor models are also shown, which are same as in Fig. \ref{fig:errorCl}.
Planck observational data in higher $\ell$ region ($\ell \simg 1000$)
provide stronger results of the parameter constraints.}
\end{figure}

According to the scalar-tensor $\Lambda{\rm CDM}$ model, the
gravitational constant has deviated by less than $1.15\%$ between
the recombination epoch and the present day at $99.99\%$ C.L..
In comparison, Nagata {\it et al.} \cite{NCS2004} and Li {\it et al.} \cite{Li} reported $G_{\rm rec}/G_0 < 1.23$ at 4$\sigma$ level
and $G_{\rm rec}/G_0 < 1.029$ at 1$\sigma$ level, respectively.
Our study places the strongest constraint on the deviation of the gravitational constant.
The CMB temperature anisotropy spectra obtained by Planck \cite{Planck}
and WMAP \cite{WMAP} are compared in Fig. \ref{fig:compare}.
Because the difference between the scalar-tensor and the $\Lambda{\rm CDM}$ models mainly arises in the high-$\ell$ region,
the observational data in the higher-$\ell$ region provide stronger
constraints on the parameters.  Therefore, the constraints from the
Planck data are much stronger than those from the WMAP ones.

\begin{figure}[ht]
\includegraphics[width=9cm,]{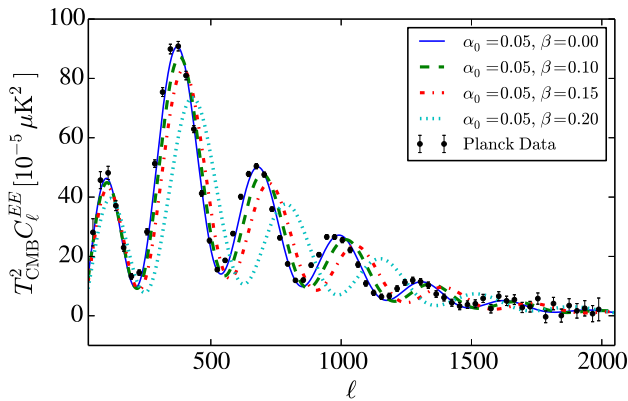}
\caption{\label{fig:errorEE} Typical CMB polarization spectra ({\it EE}) in the scalar-tensor models with the $\Lambda {\rm CDM}$ parameters
fixed to the standard values.
The data points with error bars show the Planck data.}
\end{figure}

\begin{figure}[ht]
\includegraphics[width=9cm,]{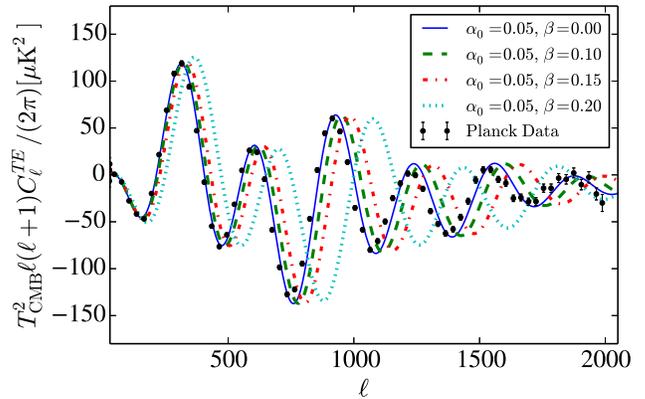}
\caption{\label{fig:errorTE} Same as Fig. \ref{fig:errorEE}, but for the temperature polarization cross spectrum.}
\end{figure}

The strong constraint in this work is also attributed to the precise
polarization spectra in the Planck 2015 results.  The scalar-tensor
model will affect the polarization spectra as well as the temperature
spectrum. The peak locations shift to the smaller scales (higher
$\ell$) and their amplitudes are suppressed.  Figures \ref{fig:errorEE} and
\ref{fig:errorTE} show the typical {\it EE} and {\it TE} CMB
polarization spectra, respectively, in the scalar-tensor $\Lambda {\rm
CDM}$ models using the same parameters as in Fig. \ref{fig:errorCl},
along with the Planck 2015 data.  Clearly the current Planck
polarization data enable us to constrain the scalar-tensor models with comparable statistical power to the temperature data.

\subsection{\label{sec:level2}Nonflat universe case}

We also performe a MCMC analysis including spatial curvature parameter $\Omega_{\rm K}$.
This is motivated by the fact that the attractor model used in this paper would predict larger gravitational constant in the past,
pushing the acoustic peaks toward smaller angular scales.
This effect could be compensated with the positive curvature which brings back the peaks toward larger angles (Nagata {\it et al.} \cite{NCS2004}).
This degeneracy, however, should be broken using the CMB data on diffusion damping scales,
because the curvature does not affect the diffusion damping whereas the variation of the gravitational constant does as discussed above.

\begin{figure}[ht] 
\includegraphics[width=7cm,]{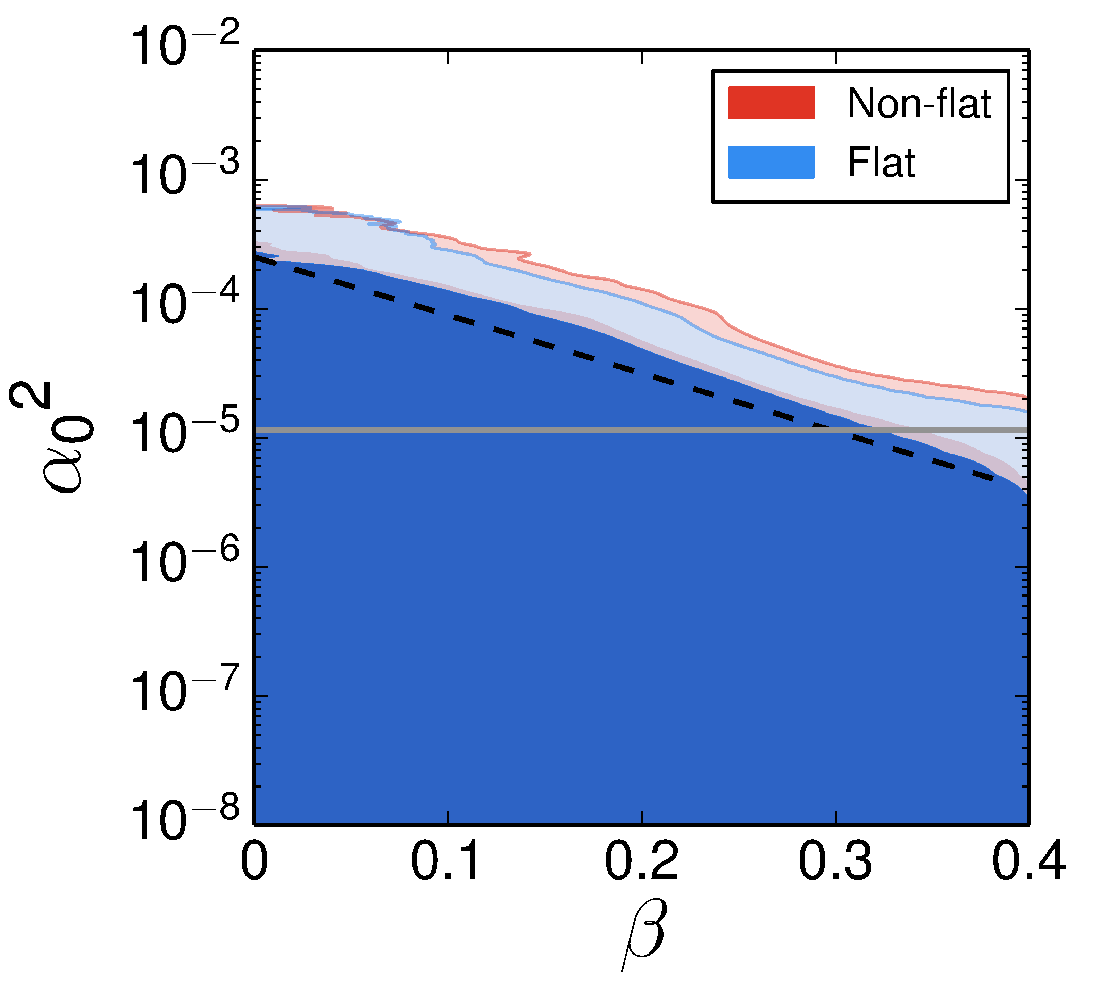}
\caption{\label{fig:compare_contour} $95.45\%$ and $99.99\%$ confidence contours in the ${\alpha_0}^2-\beta$ plane
for the scalar-tensor nonflat $\Lambda {\rm CDM}$ models with the other parameters marginalized (red),
comparing with those of the flat universe case (blue).
The black dashed line and the gray solid line show the function ${\alpha_0}^2 = 2.5 \times 10^{-4-4.5\beta}$
and the bound from the Solar System experiment, respectively.
}
\end{figure}

The constraints on the parameters ${\alpha_0}^2$ and $\beta$ in nonflat models are shown in Fig. \ref{fig:compare_contour},
where the other parameters including $\Omega_{\rm K}$ are marginalized.
We find that the constraints on the scalar-tensor coupling parameters are hardly affected by the inclusion of the spatial curvature.
This is because the angular power spectrum on small angular scales obtained from Planck is 
so precise as to break the degeneracy between the effects of varying gravitational constant and the spatial curvature.
We find that $\alpha_0^2$ is constrained as
\begin{align}
{\alpha_0}^2 &< 2.5 \times 10^{-4-4.5\beta}\ \ (95.45\%), \\
{\alpha_0}^2 &< 6.3 \times 10^{-4-4.5\beta}\ \ (99.99\%).
\end{align}
and the coupling parameter $\omega$ as
\begin{align}
\omega &> 2000\ \ (95.45\%), \\
\omega &> 790\ \ (99.99\%).
\end{align}

\begin{figure}[ht] 
\includegraphics[width=7cm,]{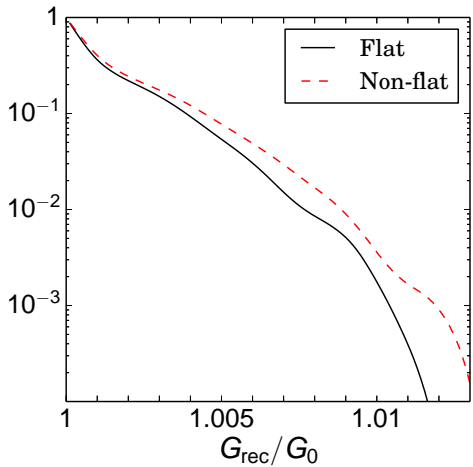}
\caption{\label{fig:compare_Gphi_1d} Posterior distribution of $G_{\rm rec}/G_0$ for the nonflat models (red dashed) compared with the flat model (black).
}
\end{figure}

Also we find that $G_{\rm rec}/G_0$ in the nonflat universe is constrained as
\begin{align}
G_{\rm rec}/G_0 &< 1.0062\ \ (95.45\%), \\
G_{\rm rec}/G_0 &< 1.0125\ \ (99.99\%).
\end{align}
The posterior distribution of $G_{\rm rec}/G_0$ is shown in Fig. \ref{fig:compare_Gphi_1d}.

Although there are a few changes in the constraints of $G_{\rm rec}/G_0$, comparing with those of the flat case,
this is in the standard deviation.
Table \ref{tab:table1} shows $68.27\%$ confidence limits of the cosmological parameters
in the scalar-tensor nonflat $\Lambda \rm CDM$ model.
These parameters are also still consistent with the those of the Planck results \cite{Planck}.
The limits on the ${\rm log}_{10}({\alpha_0}^2)$ and $\beta$ are summarized in Table \ref{tab:table2}.

These strong constraints in the nonflat universe are attributed to the lensing potential power spectrum in the Planck 2015 results.
Using the lensing potential reconstruction data leads to a strong constraint on $\Omega_{\rm K}$ \cite{Planck_lens}
and it breaks the degeneracy between the effects of varying gravitational constant
and the spatial curvature mentioned above further.
If we do not include the CMB lensing data in the scalar-tensor nonflat $\Lambda \rm CDM$ model,
we find that the variation of the gravitational constant is constrained as $G_{\rm rec}/G_0 < 1.0148$ ($99.99\%$ C.L.),
which is much weaker than the result shown above.

\section{Summary}

We have constrained the scalar-tensor $\Lambda \rm CDM$ model from the Planck data by using the MCMC method.
We have found that the present-day deviation from the Einstein gravity (${\alpha_0}^2$) is smaller than $2.5\times10^{-4-4.5\beta}$ ($95.45\%$ C.L.) and $6.3\times10^{-4-4.5\beta}$ ($99.99\%$ C.L.) for $0<\beta<0.4$.
The variation of the gravitational constant is also constrained as 
$G_{\rm rec}/G_0 < 1.0056$ ($95.45\%$ C.L.) 
and $G_{\rm rec}/G_0 < 1.0115$  ($99.99\%$ C.L.).
The significant improvement of these constraints over the previous works is attributed to 
the precise measurements of the diffusion damping effect in the temperature and the new polarization power spectra obtained by Planck.
The deviation of the gravitational constant between the recombination and 
the present epochs is found to be less than $1.15\%$ at $99.99\%$. 
We have also found that these constraints are fairly robust against the inclusion of the spatial curvature. 

\begin{table}[ht]
\caption{\label{tab:table1}
$68.27\%$ confidence limits for the standard cosmological parameters in the scalar-tensor $\Lambda \rm CDM$ model.
}
\begin{ruledtabular}
\begin{tabular}{lccr}
&\multicolumn{2}{c}{$68.27\%$\ limits}\\
\textrm{Parameter}&
\textrm{$\Omega_{\rm K} = 0$}&
\textrm{$\Omega_{\rm K} \neq 0$}\\
\colrule
$\Omega_{\rm b}h^2$ & $0.02224\pm 0.00016$ & $0.02225\pm 0.00015$\\
$\Omega_{\rm c}h^2$ & $0.1189\pm 0.0014$ & $0.1188\pm 0.0014$\\
$H_0$ & $67.92\pm 0.76$ & $66.31\pm 4.1$\\
$\tau_{\rm reio}$ & $0.069\pm 0.013$ & $0.065\pm 0.014$\\
${\rm ln}(10^{10}A_{\rm s})$ & $3.068\pm 0.024$ & $3.061\pm 0.028$\\
$n_{\rm s}$ & $0.9668\pm 0.0051$ & $0.9672\pm 0.0051$\\
\colrule
$\Omega_{\rm K}$ & --- & $-0.0046\pm 0.0096$
\end{tabular}
\end{ruledtabular}
\end{table}

\begin{table}[ht]
\caption{\label{tab:table2}
$95.45\%$ confidence limits for ${\rm log}_{10}({\alpha_0}^2)$ and $\beta$.
}
\begin{ruledtabular}
\begin{tabular}{lccr}
&\multicolumn{2}{c}{$95.45\%$\ limits}\\
\textrm{Parameter}&
\textrm{$\Omega_{\rm K} = 0$}&
\textrm{$\Omega_{\rm K} \neq 0$}\\
\colrule
${\rm log}_{10}({\alpha_0}^2)$ & $<-3.72$ & $<-3.68$\\
$\beta$ & $<3.15$ & $<3.16$\\
\end{tabular}
\end{ruledtabular}
\end{table}

\begin{acknowledgments}
This work is supported in part by MEXT Grants-in-Aid for Scientific Research on Innovative Areas, 
No. 15H05890 (N.\,S. and K.\,I.) and No. 15H05894 (T.\,C.).  
This work is also supported by 
Grants-in-Aid for Scientific Research from JSPS [No. 24540287 (T.\,C.), 24340048 (K.\,I.) and 25287057 (N.\,S.)],  and 
in part by Nihon University (T.\,C.).   

\end{acknowledgments}


\end{document}